\documentclass[11pt]{article}
\usepackage[utf8]{inputenc}
\usepackage[T1]{fontenc}
\IfFileExists{lmodern.sty}{\usepackage{lmodern}\usepackage{microtype}}{}
\usepackage[margin=1in]{geometry}
\usepackage{graphicx}
\usepackage{amsmath,amssymb}
\usepackage{booktabs}
\usepackage{array}
\usepackage{caption}
\usepackage{float}
\usepackage{authblk}
\usepackage[hidelinks]{hyperref}
\usepackage{xurl}
\usepackage{marvosym}   
\makeatletter
\renewcommand{\@fnsymbol}[1]{\Letter}   
\makeatother
\title{\textbf{Average Rankings Mask Per-Subject Optimality: A Friedman--Nemenyi Benchmark of EEG Motor-Imagery BCI Decoders}}
\author[1,2]{Xavier Vasques}
\author[3,4,5]{Paul Barbaste}
\author[3,6,7,{\protect\thanks{Corresponding author: \texttt{olivier.oullier@mbzuai.ac.ae}}}]{Olivier Oullier}
\affil[1]{IBM Technology, Bois-Colombes, France}
\affil[2]{IBM France Lab, Orsay, France}
\affil[3]{Inclusive Brains, Marseille, France}
\affil[4]{Wavestone, Paris, France}
\affil[5]{Human Technology Foundation, Paris, France}
\affil[6]{Computing and Mathematical Sciences Division, Mohamed bin Zayed University of Artificial Intelligence (MBZUAI), Abu Dhabi, UAE}
\affil[7]{Institute for Artificial Intelligence, Biotech Dental Group, Salon-de-Provence, France}
\date{}

\begin{document}
\maketitle

\begin{abstract}
Electroencephalography (EEG) is the dominant non-invasive modality for brain-computer interfaces (BCIs), yet reliable decoding of motor imagery is hampered by pronounced inter- and intra-individual variability. A recurring claim in the literature is that one decoding pipeline, most often a spatial or Riemannian method, is broadly preferable. We test the weakest version of that claim under the most favourable possible conditions. Using the Mother of All BCI Benchmarks (MOABB) framework, we evaluated 1,056 distinct decoding configurations (feature extractor $\times$ scaler $\times$ classifier), amounting to more than 340,000 subject-level model fits, across three public left-versus-right motor-imagery datasets (PhysionetMI, 109 participants; Cho2017, 52; Zhou2016, 4) and two frequency bands (8--15 Hz, 8--30 Hz). Crucially, every model is fit and tested within a single recording session of a single participant, the easiest regime, which gives every pipeline its best chance. We then apply the inferential statistics that are standard for multi-classifier comparison: Friedman omnibus tests, Nemenyi critical-difference analysis, and Wilcoxon signed-rank tests with effect sizes. Covariance tangent-space projection (cov-tgsp) and Common Spatial Patterns (CSP) form the strongest families, but their ordering is dataset-dependent and, on the largest and most heterogeneous cohort (PhysionetMI), statistically indistinguishable (Nemenyi p = 0.27; Kendall's W = 0.11). At the individual level the single best pipeline is optimal for only 35\% of PhysionetMI participants, and nonlinear descriptors are best for roughly one third of them; matching pipeline to participant would add about seven accuracy points over the best fixed choice. We show that this ranking is not an artefact of feature dimensionality, and that classifier and scaler choices are secondary to the feature representation. Even in the easiest evaluation regime, then, no single pipeline dominates: this is a lower bound on the personalization problem and a quantitative case for participant-aware model selection rather than a universal decoder.
\end{abstract}


\section{Introduction}

Brain-computer interfaces (BCIs) translate neural activity into commands for external devices, and have become a focus of both clinical and consumer neurotechnology [1,2]. For people living with severe motor or speech impairment, locked-in syndrome, amyotrophic lateral sclerosis, spinal-cord injury, or post-stroke disorders, BCIs offer a route to communication and control that no longer depends on residual movement [3--8]. Among non-invasive options, electroencephalography (EEG) remains the most widely deployed modality because it is safe, affordable, portable, and offers high temporal resolution [2].

Decoding motor imagery, the mental simulation of movement, which engages sensorimotor dynamics overlapping those of overt movement [9], is nonetheless difficult. EEG is intrinsically variable and noisy [10,11], and varies markedly both between individuals and across sessions of the same individual, driven by attention, fatigue, electrode placement, impedance, and physiology [12,13]. This non-stationarity is the central obstacle to dependable BCIs outside the laboratory [3]. It also underlies the phenomenon often labelled ``BCI illiteracy'', the observation that a substantial minority of users do not achieve reliable control [14--16].

A large methodological literature seeks features and classifiers that are robust to this variability. Spatially informed representations dominate motor-imagery decoding: Common Spatial Patterns (CSP) [17] and Riemannian-geometry methods that operate on trial covariance matrices, typically through tangent-space projection [18--20]. These often perform strongly, but are widely reported to degrade when inter-subject variability or recording conditions shift [21]. This has motivated broader feature inventories: time-, frequency-, and decomposition-domain summaries [22], and, to capture aspects of EEG complexity tied to fluctuating cognitive state, nonlinear descriptors such as Hjorth parameters, Higuchi's fractal dimension (HFD), the Hurst exponent, and singular-value-decomposition (SVD) entropy [11,23,24], as well as functional-connectivity features [25--27].

This abundance of methods raises a deceptively simple question: which pipeline should a practitioner use? The literature frequently answers with a single recommendation, most often a Riemannian method [28,29]. We argue that this framing skips a logically prior question. Before asking how well a method generalizes across users, the hard problem, which demands participant-independent protocols such as leave-one-subject-out (LOSO) cross-validation [30], one should ask whether a single best pipeline even exists under the easiest possible conditions: when a separate model is fit and tested within one session of one participant. Within-session evaluation cannot, and is not intended to, establish cross-subject transfer; it is the most forgiving regime, giving every method its best chance on every individual. If, even here, the best pipeline differs from person to person, then a universal decoder is untenable a fortiori, and the field's effort is better spent on individualization than on the search for a single winner. Establishing that bound is the goal of this study.

\section{Methods}

\subsection{Datasets}

We used three publicly available datasets through MOABB [34,35] under a single harmonized paradigm. PhysionetMI [31,38] comprises 109 participants recorded with 64 channels at 160 Hz; from the original protocol we retained the left- versus right-hand motor-imagery condition. Cho2017 [11] comprises 52 participants recorded with 64 channels at 512 Hz on a BioSemi ActiveTwo system, designed around left- and right-hand motor imagery. Zhou2016 [33] comprises only 4 participants recorded with 14 channels at 250 Hz across sessions separated by days to months; we retained the left/right condition. Because four participants cannot support meaningful population inference, Zhou2016 is reported for completeness and illustration only, is excluded from all inferential statistics, and is given no weight in our conclusions. All analyses used the MOABB LeftRightImagery paradigm, harmonizing the task to a two-class problem (left vs right hand). Native sampling rates were preserved (160/512/250 Hz). This common configuration enables direct comparison across datasets that differ in cohort size and acquisition.

\subsection{Signal processing}

EEG was band-pass filtered in two ranges standard for motor imagery: 8--15 Hz, isolating the mu rhythm, and 8--30 Hz, capturing the broader sensorimotor band including beta activity. Both bands are reported throughout; we treat the band as a factor rather than a contribution, since the choice of these two ranges is routine. Epochs were extracted over a post-cue window of 0.6--2.0 s, chosen to capture the sustained motor-imagery period while excluding the earliest post-cue evoked transients.

We deliberately kept preprocessing minimal, band-pass filtering only, with no artefact rejection, independent-component cleaning, or bad-channel interpolation, so that pipelines are compared under identical, fully reproducible inputs. This choice has an important consequence that we return to in the Discussion and Limitations: spectral and nonlinear descriptors (HFD, SVD entropy, Hurst) are more sensitive to muscular and ocular artefacts than covariance- or CSP-based features are, so any gap between the two families may partly reflect differential artefact sensitivity rather than neural information alone. We accordingly treat nonlinear results as hypotheses about individual heterogeneity to be confirmed under explicit artefact handling, not as established neural effects. The chosen window provides 224 samples at 160 Hz, which is short for estimators of long-range temporal structure such as the Hurst exponent and SVD entropy; the stability of these estimators at this length is itself uncertain and is one reason we do not over-interpret their absolute values.

\subsection{Feature extraction}

Four methodologically distinct feature families were evaluated. (i) CSP [17,39] derived low-dimensional spatial filters maximizing inter-class variance differences, computed separately within each band (no filter-bank CSP); we used two components, i.e. two features per trial. (ii) Covariance tangent-space projection (cov-tgsp) [19,40]: per-epoch channel covariance matrices were projected into the tangent space at their Riemannian geometric mean, yielding Euclidean vectors that preserve covariance structure. (iii) Coherence-based connectivity projected into tangent space (con-tgsp) [25], summarizing pairwise channel interactions in a Euclidean-compatible form. (iv) Nonlinear descriptors, HFD (with maximum scale k = 10), Hjorth mobility and complexity [23], SVD entropy, and the Hurst exponent, computed per channel and concatenated.

Feature dimensionality differs substantially across families: 2 features for CSP, 64 for HFD and SVD entropy, up to 192 for Hjorth, and up to 2,080 for the two tangent-space methods (for 64 channels). This is intrinsic to the methods and is not a nuisance we removed; instead we analyse its relationship to performance directly, because a fair reading of any cross-family comparison must ask whether ranking simply tracks dimensionality. The compared pipelines should therefore be read as complete end-to-end decoding configurations, not as dimension-matched representations.

\subsection{Scaling, classifiers, and the configuration count}

Each feature representation was passed through one of four scalers (StandardScaler, RobustScaler, MinMaxScaler, or L2 normalization) and one of several classifiers: logistic regression with elastic-net regularization, linear discriminant analysis, linear and RBF support-vector machines, random forests, and a predefined family of 17 multilayer perceptrons differing in depth, width, activation, solver, and regularization. Classifier hyperparameters were fixed within each configuration; classifiers are thus treated as predefined pipeline components, not as the product of a separate tuning stage. Random states for stochastic optimizers were fixed (seed 42). How much these scaler and classifier choices matter, relative to the choice of feature family, is examined.

The full grid comprises 1,056 distinct pipeline configurations per band and dataset (feature extractor $\times$ scaler $\times$ classifier: 12 feature extractors $\times$ 4 scalers $\times$ 22 classifiers). We state this explicitly to avoid the inflation that an aggregate ``number of evaluations'' invites: prior framings have headlined the product of configurations, subjects, bands, and datasets ($\approx$342,604), but most of those are repeated evaluations of the same 1,056 configurations on different individuals, not distinct scientific hypotheses. We therefore report the number of distinct configurations (1,056) separately from the number of subject-level model fits they generate (>340,000), and we de-emphasize the aggregate as a headline (Fig. 1). A breakdown is given in Table 1. The complete grid is summarized at the family level in the main analyses; the per-subject inferential comparison uses a representative logistic-regression pipeline per feature family, the configuration where the strongest performance consistently concentrated.

\begin{figure}[!htbp]\centering
\includegraphics[width=0.96\linewidth]{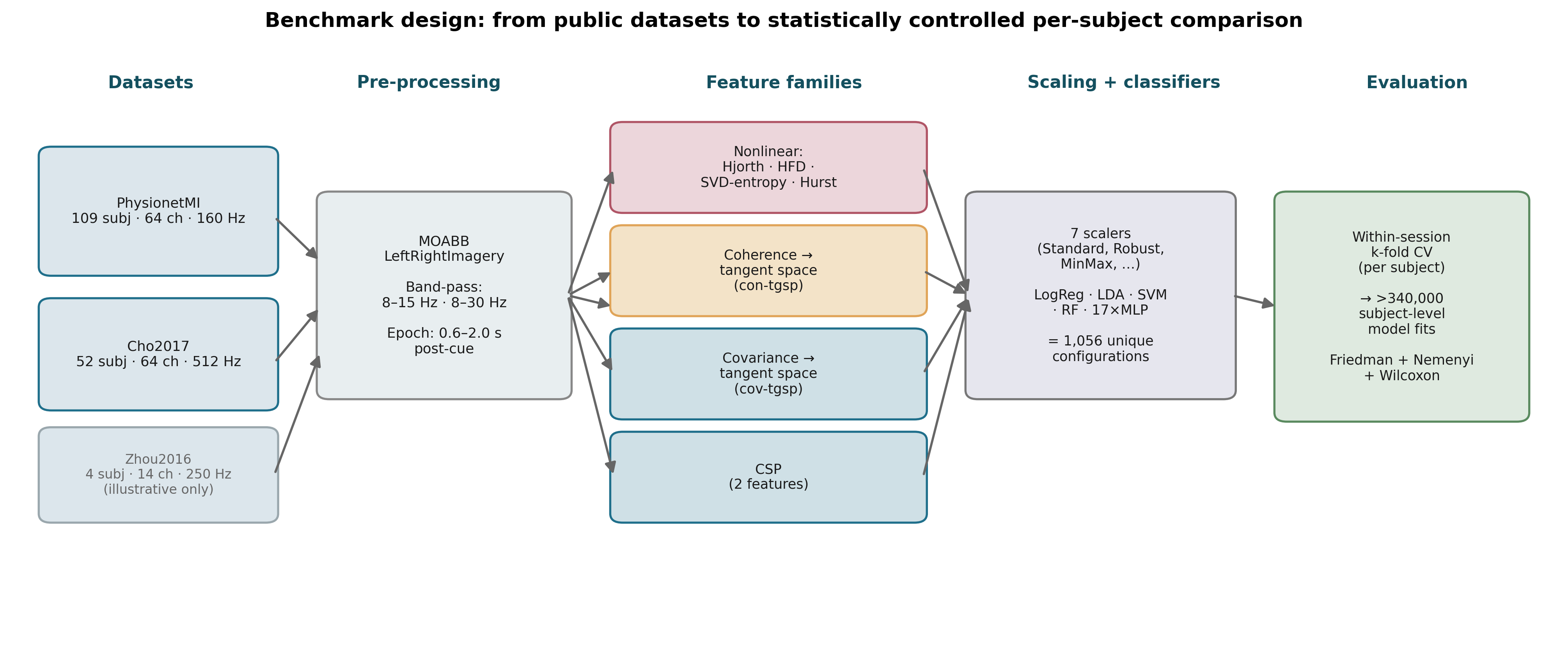}
\caption{Benchmark design. Three public motor-imagery datasets are harmonized through the MOABB LeftRightImagery paradigm, band-pass filtered (8--15 Hz and 8--30 Hz) and epoched (0.6--2.0 s post-cue). Four feature families (CSP; covariance and coherence tangent-space projections; nonlinear descriptors) are combined with four scalers and a set of classical classifiers, yielding 1,056 distinct configurations per band and dataset. Every configuration is fit and tested within a single recording session of a single participant; the resulting per-subject scores feed the statistical comparison. Zhou2016 (greyed) is illustrative only and excluded from inference.}
\end{figure}

\subsection{Evaluation and statistical analysis}

Performance was assessed with the within-session protocol of MOABB [35]: each participant's single-session data are split into training and test folds, and every pipeline is trained and tested on identical splits. This yields one accuracy per participant per pipeline per band. We use accuracy as the primary metric; because the LeftRightImagery paradigm is, by construction, a balanced two-class problem, accuracy and balanced accuracy nearly coincide here, and we report accuracy for direct comparability with MOABB and the wider literature. We emphasize that within-session evaluation measures generalization only to unseen trials of the same session; it makes no claim about cross-session or participant-independent transfer.

To move beyond means and standard deviations, we applied the inferential framework standard for comparing multiple classifiers across a sample of cases [36,37]. For each dataset (band-averaged per participant) we ran a Friedman omnibus test across the six representative pipelines, reporting Kendall's coefficient of concordance W as an effect size for the degree of agreement in the ranking. Where the omnibus test was significant we performed Nemenyi post-hoc comparisons and summarized them as critical-difference (CD) diagrams. The headline pairwise contrast (cov-tgsp vs CSP) was additionally tested with a Wilcoxon signed-rank test, with the rank-biserial correlation as effect size. We also computed 95\% confidence intervals for each pipeline's mean accuracy. Finally, for each participant we identified the single best pipeline (``winner-per-subject'') and the per-subject oracle accuracy (the accuracy obtained if the best pipeline were chosen for each individual), and contrasted the latter with the best single fixed pipeline. The same Friedman framework was applied, on the largest cohort and within a fixed feature family, to assess the influence of classifier and scaler choice. All inference was restricted to PhysionetMI and Cho2017; Zhou2016 (n = 4) was excluded.

\section{Results}

\subsection{Two families lead, but their order depends on the dataset}

Across all datasets and bands, two feature families separated clearly from the rest. Covariance tangent-space projection (cov-tgsp) achieved the highest overall mean accuracy (0.687), followed by CSP (0.651); the connectivity and nonlinear families trailed (con-tgsp 0.599; Hjorth 0.567; SVD entropy 0.563; HFD 0.555). This reproduces the familiar finding that spatial and Riemannian methods are strong default choices for motor imagery.

The ordering within the leading pair was, however, dataset-dependent (Table 2; Fig. 2). On Cho2017, cov-tgsp led decisively (0.740, 95\% CI 0.706--0.775) over CSP (0.677, 0.637--0.717). On PhysionetMI, the largest and most heterogeneous cohort, the two were close (cov-tgsp 0.644, 0.613--0.675; CSP 0.615, 0.583--0.647), with broadly overlapping per-subject distributions. On Zhou2016 the order even reversed, CSP (0.869) edging cov-tgsp (0.841). The per-subject score distributions (Fig. 2) make the overlap visible: on PhysionetMI the spatial and nonlinear families overlap substantially, whereas on Cho2017 the separation is wider.

\begin{table}[!htbp]\centering
\caption{Decomposition of the benchmark size.}
\small
\begin{tabular}{l r >{\raggedright\arraybackslash}p{7.6cm}}
\toprule
\textbf{Quantity} & \textbf{Count} & \textbf{What it represents} \\
\midrule
Distinct pipeline configurations & 1,056 & feature extractor $\times$ scaler $\times$ classifier, per band and dataset; the distinct scientific hypotheses \\
Focused subject-level evaluations & 2,076 & representative LR pipelines $\times$ participants $\times$ datasets $\times$ bands, used in the inferential analysis \\
Subject-level model fits (full grid) & > 340,000 & every configuration fit within each participant's session, across cohorts and bands \\
\bottomrule
\end{tabular}
\end{table}

\begin{table}[!htbp]\centering
\caption{Mean within-session accuracy (logistic-regression pipelines) per feature family and dataset, band-averaged, with 95\% confidence intervals for the two inferentially analysed cohorts. Zhou2016 (n = 4) is shown without intervals and excluded from inference.}
\small
\begin{tabular}{l l c c c}
\toprule
\textbf{Pipeline} & \textbf{Family} & \textbf{PhysionetMI (n=109)} & \textbf{Cho2017 (n=52)} & \textbf{Zhou2016 (n=4)} \\
\midrule
cov-tgsp & Spatial/Riemannian & 0.644 [0.613, 0.675] & 0.740 [0.706, 0.775] & 0.841 \\
CSP & Spatial/Riemannian & 0.615 [0.583, 0.647] & 0.677 [0.637, 0.717] & 0.869 \\
con-tgsp & Connectivity & 0.542 [0.519, 0.566] & 0.685 [0.652, 0.717] & 0.737 \\
Hjorth & Nonlinear & 0.547 [0.528, 0.567] & 0.584 [0.562, 0.605] & 0.672 \\
HFD & Nonlinear & 0.540 [0.519, 0.561] & 0.563 [0.547, 0.578] & 0.668 \\
SVD-En & Nonlinear & 0.539 [0.517, 0.561] & 0.587 [0.564, 0.610] & 0.682 \\
\bottomrule
\end{tabular}
\end{table}

\begin{figure}[!htbp]\centering
\includegraphics[width=\linewidth]{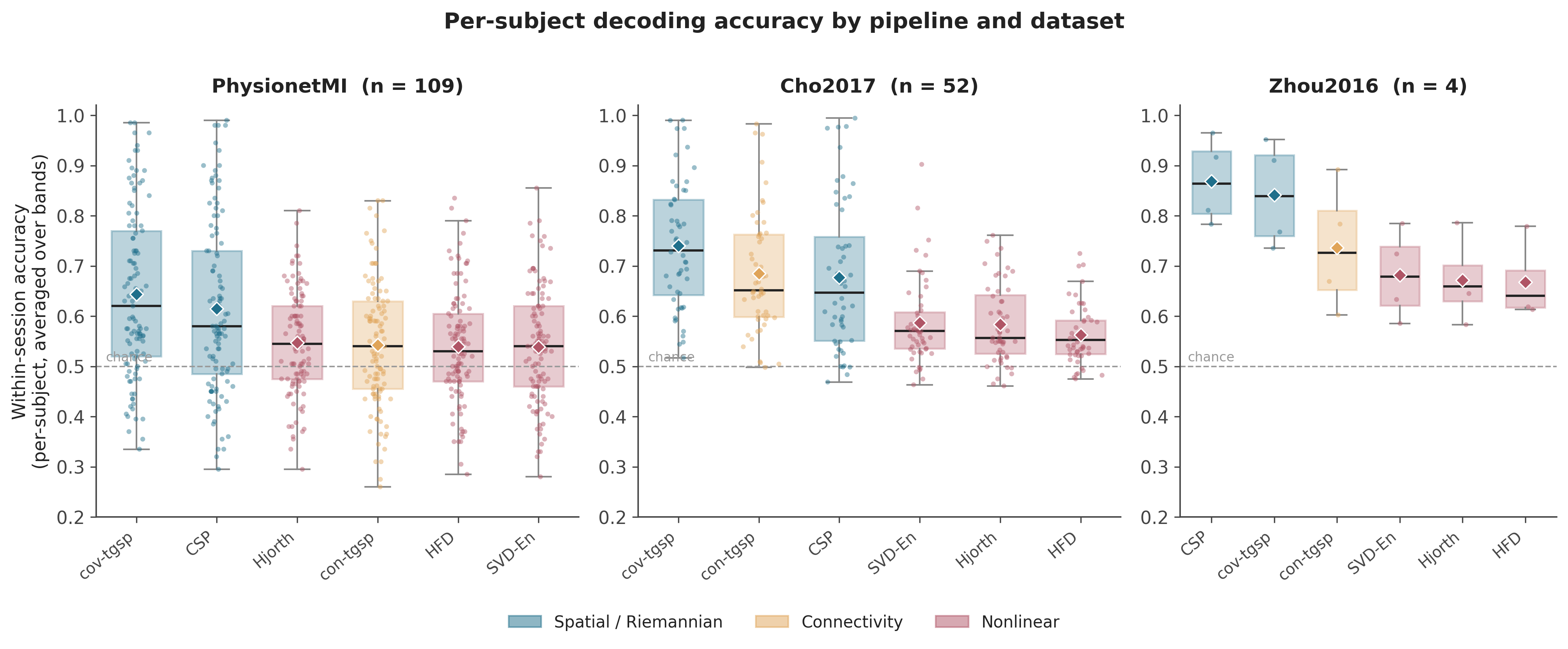}
\caption{Per-subject decoding accuracy by pipeline and dataset. Boxplots show the distribution of per-participant within-session accuracy (band-averaged); points are individual participants; diamonds mark the family mean; the dashed line is chance (0.50). Pipelines are coloured by family. On PhysionetMI (left) the leading spatial pipelines overlap heavily with one another and with the nonlinear families; on Cho2017 (centre) the separation is clearer. Zhou2016 (right, n = 4) is illustrative only.}
\end{figure}

\subsection{The leading pipelines are statistically tied on the most heterogeneous cohort}

Descriptive ranking is not evidence of an ordering. We therefore tested the six representative pipelines per dataset (Fig. 3). On both cohorts the Friedman omnibus test was significant (PhysionetMI $\chi^{2}$ = 58.3, p = 2.7$\times$10\textsuperscript{-11}; Cho2017 $\chi^{2}$ = 132.9, p = 5.9$\times$10\textsuperscript{-27}), confirming that not all pipelines perform equally. But the effect sizes tell sharply different stories. On Cho2017, concordance was moderate-to-strong (Kendall's W = 0.51): cov-tgsp occupied a clean top rank (mean rank 1.42), significantly ahead of CSP by both Nemenyi (p = 1$\times$10\textsuperscript{-4}) and Wilcoxon (p = 3.1$\times$10\textsuperscript{-6}; rank-biserial r = 0.58; cov-tgsp best for 41 of 52 participants). On PhysionetMI, by contrast, concordance was weak (W = 0.11): although cov-tgsp held the best mean rank (2.48) ahead of CSP (3.02), the two were not separable by the Nemenyi critical difference (p = 0.27), and the Wilcoxon contrast was marginal (p = 0.021, r = 0.19; cov-tgsp best for 65 of 109 participants, a 60/40 split). In other words, on the dataset that best represents real-world heterogeneity, the field's two leading pipelines are statistically indistinguishable, and the apparent ordering reported by point estimates alone is not warranted. The critical-difference diagrams (Fig. 3) capture this directly: a wide top clique on PhysionetMI versus an isolated winner on Cho2017.

\begin{figure}[!htbp]\centering
\includegraphics[width=\linewidth]{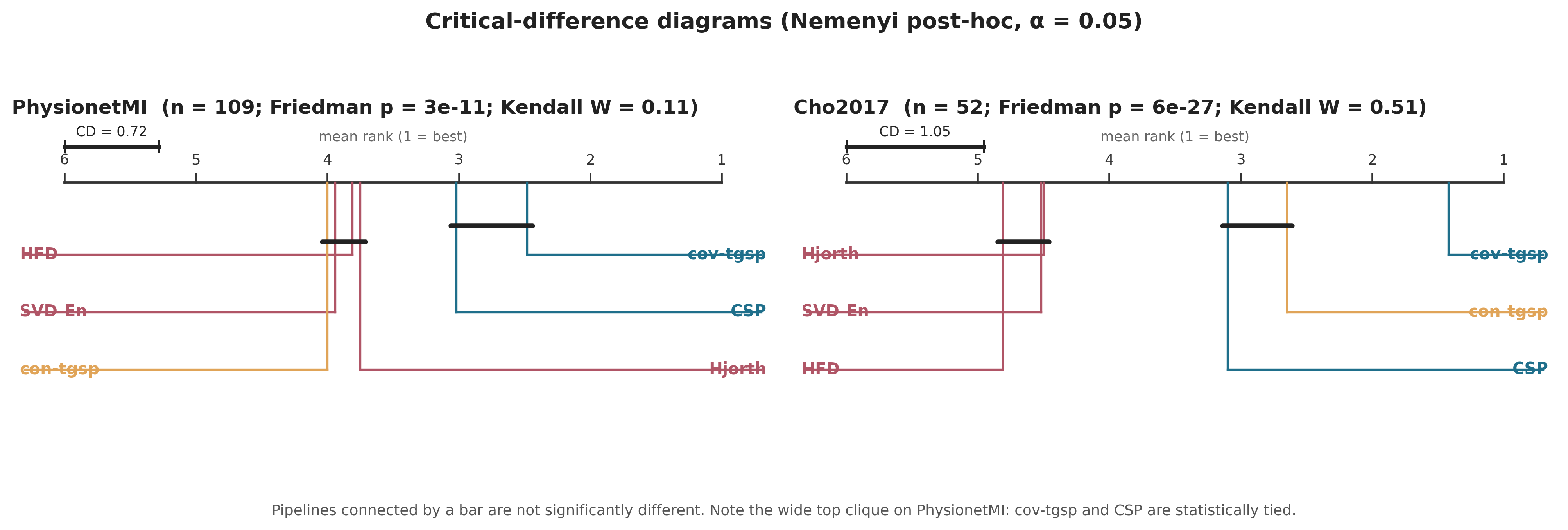}
\caption{Critical-difference diagrams (Nemenyi post-hoc, $\alpha$ = 0.05). Pipelines are placed by mean rank (1 = best); horizontal bars connect groups that are not significantly different (critical difference CD shown top-left of each panel). On PhysionetMI (left) cov-tgsp and CSP fall in a single wide top clique, statistically tied, despite the significant omnibus test and weak overall concordance (W = 0.11). On Cho2017 (right) cov-tgsp stands apart as the sole top method (W = 0.51).}
\end{figure}

\begin{table}[!htbp]\centering
\caption{Statistical comparison of the six representative pipelines, per inferentially analysed cohort. The omnibus test is significant on both datasets, but concordance (Kendall's W) and the decisive cov-tgsp-vs-CSP contrast differ markedly: the two leading pipelines are statistically tied on PhysionetMI and clearly separated on Cho2017.}
\small
\begin{tabular}{l c c}
\toprule
\textbf{Test / quantity} & \textbf{PhysionetMI (n=109)} & \textbf{Cho2017 (n=52)} \\
\midrule
Friedman $\chi^{2}$ (6 pipelines) & 58.3 & 132.8 \\
Friedman p & 2.7$\times$10\textsuperscript{-11} & 5.9$\times$10\textsuperscript{-27} \\
Kendall's W (concordance) & 0.11 (weak) & 0.51 (moderate--strong) \\
Mean rank: cov-tgsp / CSP & 2.48 / 3.02 & 1.42 / 3.10 \\
Nemenyi p (cov-tgsp vs CSP) & 0.27 (n.s.) & 1$\times$10\textsuperscript{-4} (sig.) \\
Wilcoxon p (cov-tgsp vs CSP) & 0.021 & 3.1$\times$10\textsuperscript{-6} \\
Rank-biserial r & 0.19 & 0.58 \\
cov-tgsp wins (per-subject) & 65 / 109 & 41 / 52 \\
Critical difference (CD) & 0.72 & 1.05 \\
\bottomrule
\end{tabular}
\end{table}

\subsection{At the individual level, the best pipeline varies across participants}

The tie at the top is a symptom of a deeper pattern: the identity of the best pipeline varies from participant to participant (Fig. 4). On PhysionetMI, the overall winner (cov-tgsp) was the best pipeline for only 38 of 109 participants (35\%); CSP was best for 28 (26\%); and nonlinear descriptors were best for 35 participants (32\%), with Hjorth, HFD, and SVD entropy each leading for a non-trivial share. Grouped by family, spatial/Riemannian methods won for 61\% of participants, nonlinear for 32\%, and connectivity for 7\%. On Cho2017 the picture was more concentrated but still not universal: cov-tgsp was best for 35 of 52 participants (67\%), spatial methods for 87\% overall, and the remaining 13\% were split between connectivity and nonlinear features.

This heterogeneity carries a measurable cost. Selecting the single best fixed pipeline per dataset yielded 0.644 (PhysionetMI) and 0.740 (Cho2017); matching pipeline to participant (the per-subject oracle) yielded 0.713 and 0.753, gains of about 7.0 and 1.2 accuracy points, respectively (Fig. 4, right). The gain is largest precisely where heterogeneity is largest. We report this oracle as a descriptive ceiling that bounds what participant-aware selection could recover; realizing it in practice (for instance through subject-level model selection or adaptive switching) is a separate problem that we do not address here, and is an obvious direction for future work.

\begin{figure}[!htbp]\centering
\includegraphics[width=\linewidth]{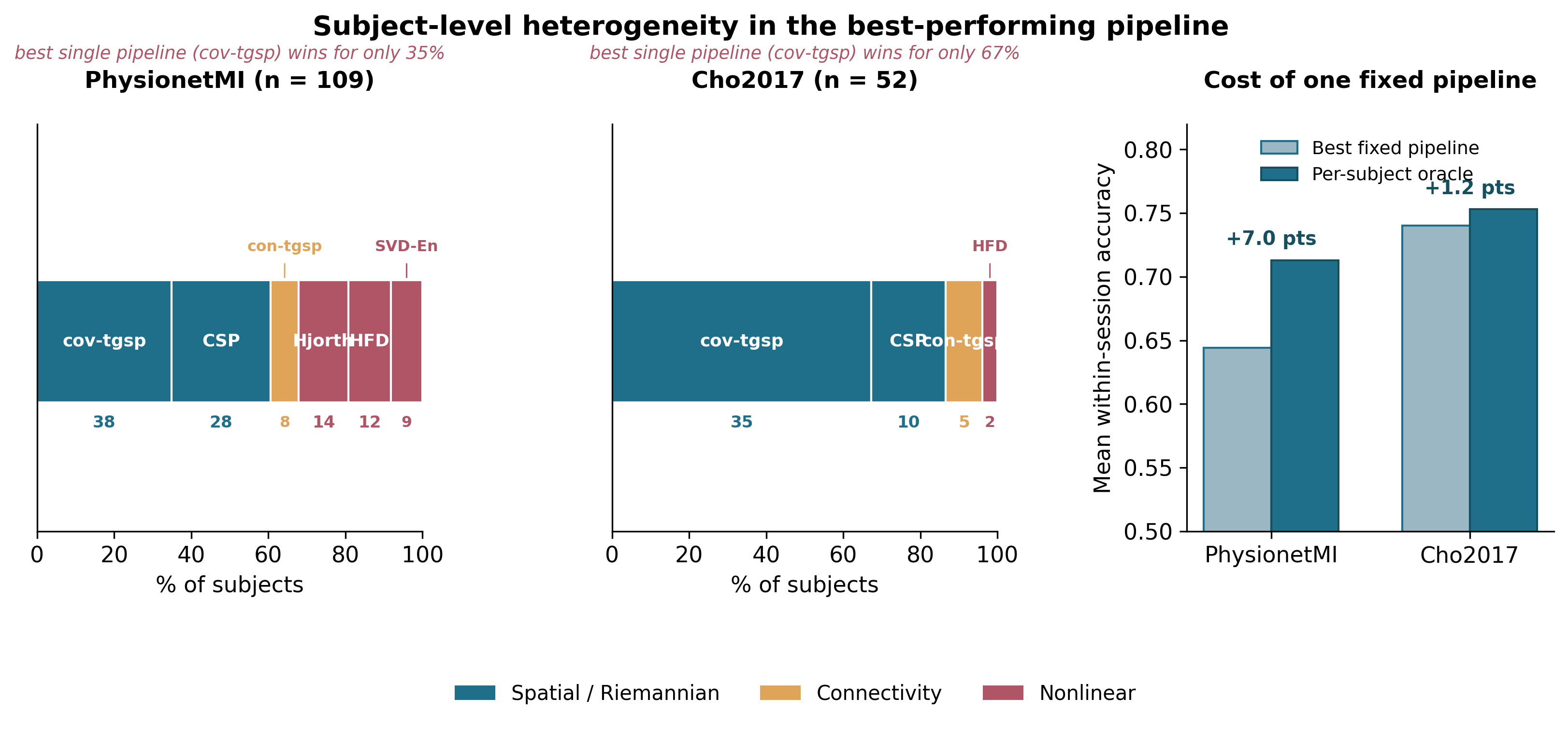}
\caption{Subject-level heterogeneity in the best-performing pipeline. Left and centre: the proportion of participants for whom each pipeline is the single best, by dataset; segment counts give the number of participants. The overall-best pipeline (cov-tgsp) is best for only 35\% of PhysionetMI participants and 67\% of Cho2017 participants. Right: the best single fixed pipeline versus the per-subject oracle; matching pipeline to participant would add about 7 accuracy points on PhysionetMI and about 1 on Cho2017.}
\end{figure}

\subsection{The ranking is not an artefact of feature dimensionality}

Because the families differ greatly in dimensionality (2 to 2,080 features) and share the same downstream classifiers, a natural concern is that the comparison merely rewards higher-dimensional representations. The data do not support this (Fig. 5). Performance is non-monotonic in dimensionality: CSP, with just 2 features, outperformed all three nonlinear families (64--192 features); and the two highest-dimensional methods, cov-tgsp and con-tgsp, share identical dimensionality (2,080) yet sit at opposite ends of the performance range (0.644 vs 0.542 on PhysionetMI). Across the six pipelines, the rank correlation between dimensionality and mean accuracy was weak and non-significant (Spearman $\rho$ = 0.26, p = 0.61 on PhysionetMI; $\rho$ = 0.53, p = 0.28 on Cho2017). What distinguishes the strong pipelines is therefore the structure of the representation, the covariance geometry exploited by cov-tgsp, rather than its size. This does not eliminate dimensionality as a consideration in any single pairwise comparison, but it rules it out as the explanation for the family-level ordering.

\begin{figure}[!htbp]\centering
\includegraphics[width=0.82\linewidth]{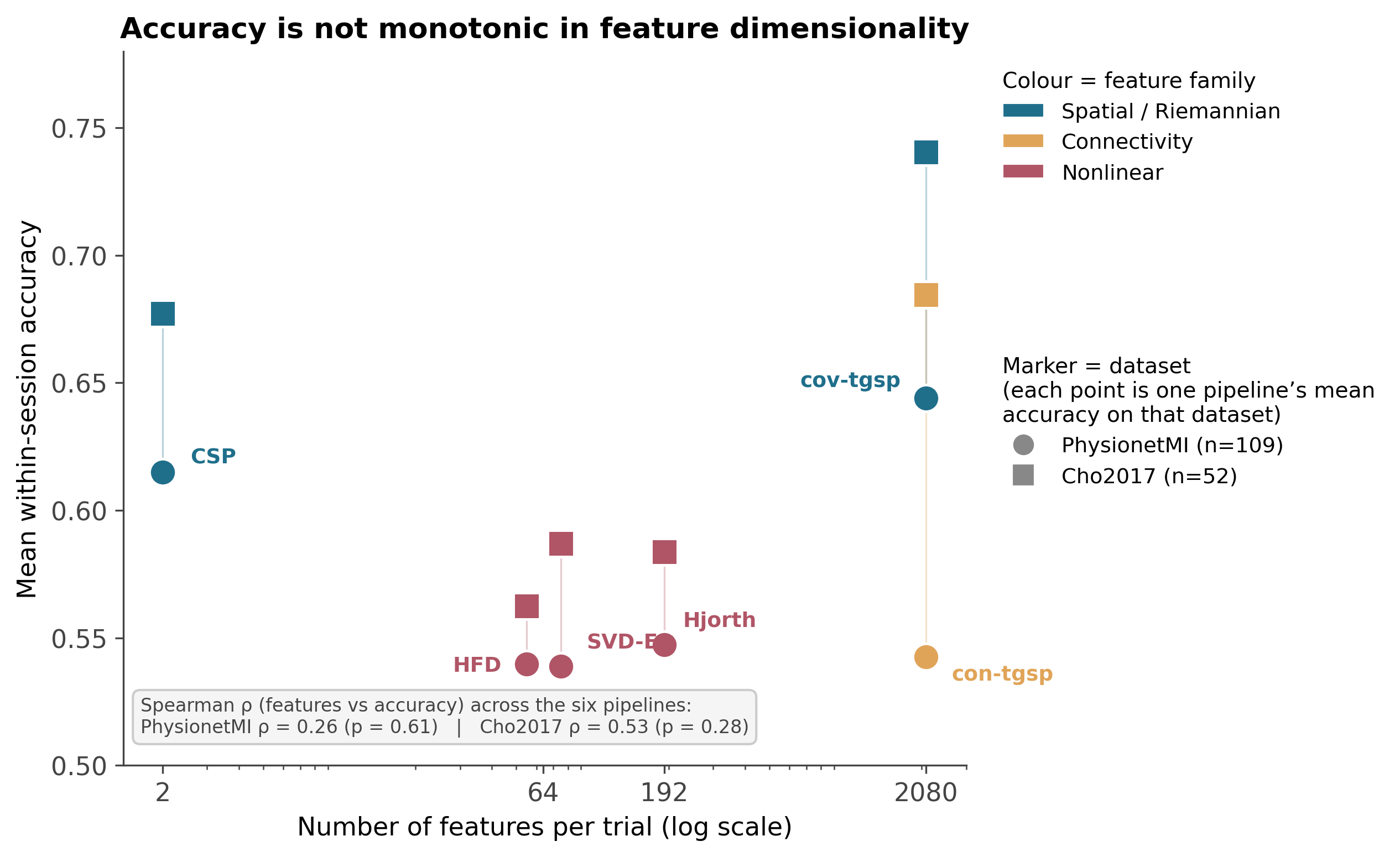}
\caption{Accuracy is not monotonic in feature dimensionality. Each point is one pipeline's mean within-session accuracy on a dataset, against the number of features per trial (log scale); colour denotes feature family and marker denotes dataset. Points sharing a dimensionality are given a small horizontal offset for legibility. CSP (2 features) beats the nonlinear families (64--192); cov-tgsp and con-tgsp share 2,080 features but differ widely in accuracy. The rank correlation between dimensionality and accuracy is weak and non-significant.}
\end{figure}

\subsection{Within a feature family, a linear classifier and a standard scaler suffice}

Each configuration also fixes a scaler and a classifier, and a natural question is how much these downstream choices matter relative to the feature family. We examined this on PhysionetMI, the largest cohort, holding the feature family fixed so that classifier and scaler effects are not confounded with the feature effect (Fig. 6; Table 4). With covariance tangent-space features, classifier choice produced a significant but modest effect (Friedman $\chi^{2}$ = 84.5, p < 10\textsuperscript{-15}; Kendall's W = 0.16). Regularized linear decision rules performed best: linear support-vector machines (mean accuracy 0.639) and logistic regression (0.628) led, followed by the RBF support-vector machine (0.613) and random forests (0.607), while the averaged multilayer perceptron (0.572) and linear discriminant analysis (0.561) trailed. Linear discriminant analysis was, by contrast, competitive on the two-dimensional CSP representation (0.610) and degraded only on the high-dimensional tangent-space features, in line with its known sensitivity to dimensionality in the absence of shrinkage. Among the four strongest classifiers the spread was about three accuracy points.

Scaler choice had an even smaller influence. On covariance tangent-space features the three standard scalers were close (MinMaxScaler 0.594, RobustScaler 0.592, StandardScaler 0.585), and only L2 normalization clearly trailed (0.550); the omnibus effect, although significant given the sample size (Friedman $\chi^{2}$ = 93.7, p < 10\textsuperscript{-19}), corresponded to a spread of at most four accuracy points and was driven almost entirely by the L2 normalizer. For CSP the scalers were closer still. These analyses indicate that, once a strong feature family is chosen, a regularized linear classifier with a standard or robust scaler is a reliable default, and the substantial per-subject heterogeneity documented above is a property of the feature representation and the individual rather than of the classifier or the scaler.

\begin{figure}[!htbp]\centering
\includegraphics[width=\linewidth]{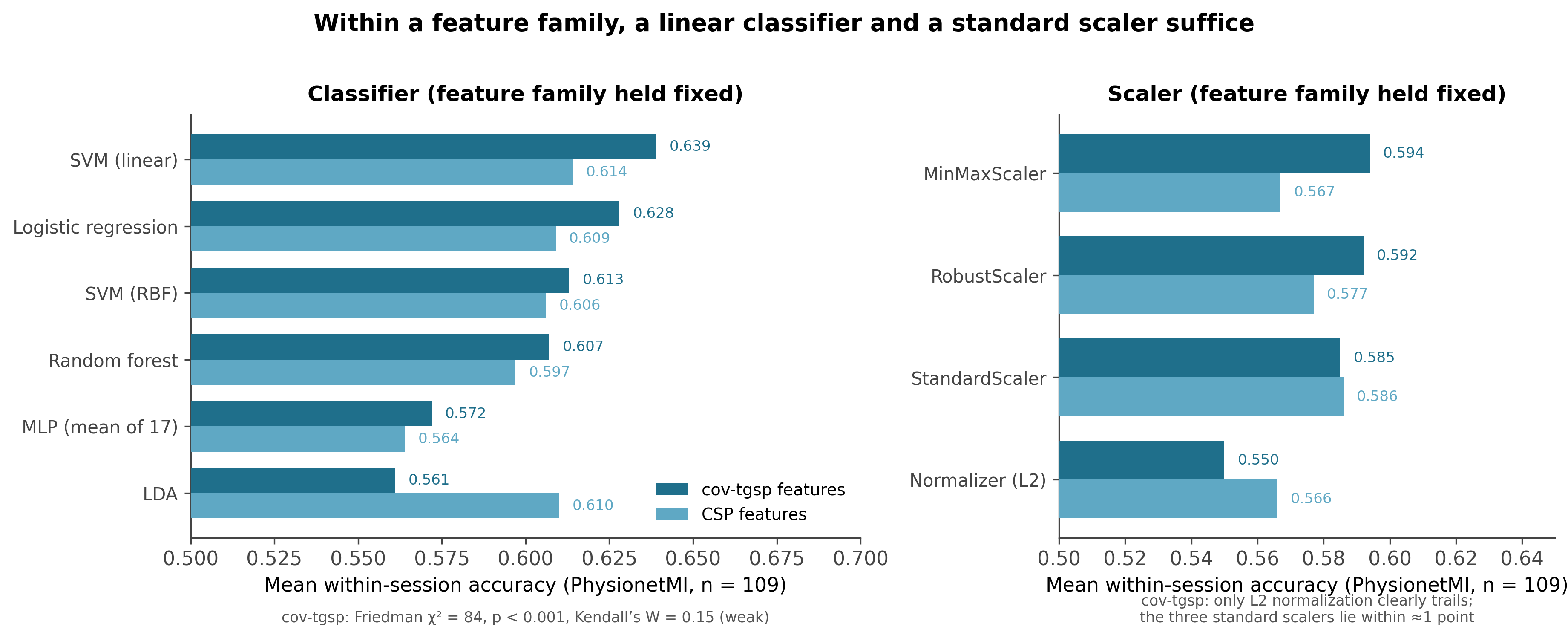}
\caption{Effect of classifier and scaler choice on PhysionetMI (n = 109), with the feature family held fixed so that the comparison is not confounded by feature type or dimensionality. Bars show mean within-session accuracy for covariance tangent-space features (dark) and CSP features (light). Left: among classifiers, regularized linear models lead on tangent-space features; linear discriminant analysis is competitive on the low-dimensional CSP representation but not on the high-dimensional tangent-space one. Right: among scalers, the three standard scalers are nearly interchangeable and only L2 normalization clearly trails.}
\end{figure}

\begin{table}[!htbp]\centering
\caption{Effect of classifier and scaler choice on PhysionetMI (n = 109), mean within-session accuracy with the feature family held fixed. Values are band-averaged per participant and then averaged across participants. The multilayer perceptron entry is the mean over the 17 predefined architectures.}
\small
\begin{tabular}{l c c}
\toprule
\textbf{Pipeline component} & \textbf{cov-tgsp features} & \textbf{CSP features} \\
\midrule
\addlinespace
\multicolumn{3}{@{}l}{\textit{Classifier (4 scalers averaged)}} \\
SVM (linear) & 0.639 & 0.614 \\
Logistic regression & 0.628 & 0.609 \\
SVM (RBF) & 0.613 & 0.606 \\
Random forest & 0.607 & 0.597 \\
MLP (mean of 17) & 0.572 & 0.564 \\
LDA & 0.561 & 0.610 \\
\addlinespace
\multicolumn{3}{@{}l}{\textit{Scaler (22 classifiers averaged)}} \\
MinMaxScaler & 0.594 & 0.567 \\
RobustScaler & 0.592 & 0.577 \\
StandardScaler & 0.585 & 0.586 \\
Normalizer (L2) & 0.550 & 0.566 \\
\bottomrule
\end{tabular}
\end{table}

\section{Discussion}

Under the most favourable evaluation regime available, within-session, per-participant model fitting, no single decoding pipeline is best for most users. Covariance tangent-space projection and CSP are the strongest families on average, consistent with the wider literature [28,29], but their advantage is conditional: it is robust on the more homogeneous Cho2017 cohort and weakens, in the strict statistical sense, on the larger and more heterogeneous PhysionetMI cohort, where the two leading methods are indistinguishable and the best pipeline differs from person to person. Because within-session evaluation is the easiest case, this is a lower bound: a method that cannot be declared universally best even here is unlikely to be universally best under the harder, more realistic cross-session and cross-subject regimes. The practical implication is not that spatial methods should be abandoned, as they remain a sound default, but rather that effort may be productively directed toward participant-aware selection in addition to the continued refinement of strong default pipelines.

The classifier and scaler analyses sharpen this into a concrete recommendation. Within a strong feature family, a regularized linear classifier was consistently among the best choices, while more flexible models (RBF support-vector machines, random forests, multilayer perceptrons) offered little or no advantage. This is consistent with the rationale of tangent-space mapping, which projects covariance matrices into a Euclidean space where the classes are approximately linearly separable [19,20,28]; once that representation is in place, additional classifier complexity has limited room to help. Scaler choice was smaller still, with standard, robust, and min-max scaling nearly interchangeable. The per-subject heterogeneity we observe therefore originates in the feature representation and the individual, not in the downstream classifier or scaler, which behave as second-order factors once a strong feature family is selected. For practitioners, this suggests that personalization effort is better spent on the choice of feature representation per participant than on classifier or scaler tuning.

This reframing also speaks to ``BCI illiteracy''. The term has often been read as a property of users, that some people simply cannot operate a BCI [14,15]. Our results are more consistent with the alternative reading, increasingly voiced in the field, that much of what is called illiteracy reflects a mismatch between an individual's neurophysiology and the fixed decoding algorithm imposed on them [16]: nonlinear descriptors were the best representation for roughly a third of PhysionetMI participants, who would be disadvantaged by a spatial-only default. ``BCI inefficiency'', in this view, may be at least partly a property of the system rather than the person, and in principle addressable by changing the pipeline rather than the user.

Why does the leading-pair ordering depend on the dataset? PhysionetMI is recorded at a lower sampling rate and is widely noted for greater protocol and signal heterogeneity than Cho2017 [31,38], and our per-subject dispersion is correspondingly larger there. The dispersion is, moreover, largely pipeline-independent, pointing to heterogeneity in the underlying signal rather than in any one method. Inter-subject EEG variability is known to exceed intra-subject variability and to reflect stable individual traits, anatomy, functional connectivity, and cortical folding, more than momentary task effects [12,13]. From a machine-learning standpoint this is covariate shift: the feature distribution that is optimal for one individual is not optimal for another, which is one reason a fixed pipeline can leave accuracy unused and why transfer-learning and domain-adaptation strategies have become central to cross-subject BCI [21,41]. Our within-session design does not engage that machinery; it establishes the prior fact that motivates it.

Framing variability as covariate shift connects our results to a substantial body of work on transfer learning and domain adaptation for BCIs, which seeks to reuse data or models across sessions and subjects despite shifting feature distributions [44]. Approaches range from Riemannian alignment of covariance matrices and subspace transfer [21] to Euclidean-space data alignment [46] and multi-task formulations that learn shared structure across users [44]. Inter- and intra-subject variability has itself been the subject of dedicated review, with broad agreement that it is a dominant determinant of BCI performance and a principal target for individualization [45]. Our contribution is upstream of these methods: by quantifying, under the easiest evaluation regime, how often the best pipeline changes from person to person, we provide a baseline against which the added value of a given adaptation strategy can be measured. A complementary line of work pursues end-to-end deep architectures such as compact convolutional networks [42] and deep and shallow convolutional networks [43], which fold feature extraction and classification into a single trainable model. These methods were outside our scope, which was deliberately restricted to interpretable classical pipelines, but applying the same per-subject, effect-size-aware analysis to them, and asking whether they reduce or merely relocate the heterogeneity we observe, is a natural extension.

Several constraints bound our conclusions and define the work needed to extend them. First and most important, the evaluation is within-session by design; it quantifies a necessary condition for universal decoding and says nothing directly about cross-session or LOSO transfer. Replicating this analysis under participant-independent protocols is the essential next step, and we expect the heterogeneity documented here to widen rather than narrow under that harder regime. Second, preprocessing was intentionally minimal. Nonlinear descriptors are more sensitive to muscular and ocular artefacts than covariance-based features are, so part of the family-level gap, and part of the per-subject success of nonlinear features, could reflect artefact sensitivity rather than neural information; the proper test is to repeat the benchmark with explicit artefact handling (for example ICA, ASR, or automated pipelines) and ask whether the winner-per-subject pattern survives. We therefore present the nonlinear results as individual-level hypotheses, not settled neural effects. Third, the 0.6--2.0 s window yields short segments (224 samples at 160 Hz) for long-memory estimators such as the Hurst exponent and SVD entropy, whose stability at this length is uncertain and likely subject-dependent; absolute values for these descriptors should be treated cautiously. Fourth, accuracy is the only reported metric; this is defensible here because the paradigm is balanced by construction, but balanced accuracy, AUROC, and calibration should accompany any extension to imbalanced paradigms. Fifth, the classifier and scaler analyses used the full grid, which was available for PhysionetMI; confirming the same pattern on other cohorts would strengthen the practical recommendation. Sixth, Zhou2016 (n = 4) cannot support inference and was used for illustration only. Finally, our comparison is confined to classical pipelines; deep architectures (e.g. EEGNet, shallow and deep ConvNets, transformers) and adaptive online methods were outside scope, and a unified comparison under the same statistical lens is a natural follow-up.

The MOABB initiative remains the reference reproducibility effort in EEG BCI, evaluating dozens of pipelines across many datasets and reporting that Riemannian methods are broadly robust [34,35]. Our study is narrower in pipeline breadth but adds two things that aggregate benchmarks typically omit: an explicit, effect-size-aware statistical treatment of the per-dataset ranking, and a participant-level decomposition that asks not which method wins on average but for how many individuals it actually wins. The two views are complementary: average robustness and per-subject heterogeneity are both real, and the second is what a deployed system must contend with.

\section{Conclusion}

A large within-session benchmark, equipped with the statistics standard for multi-classifier comparison, shows that covariance tangent-space projection and CSP are the strongest motor-imagery decoders on average but are not universally best: their ordering depends on the dataset, they are statistically tied on the largest and most heterogeneous cohort, and the single best pipeline is optimal for only about a third of participants there, with nonlinear descriptors best for roughly another third. Feature dimensionality does not explain this pattern, and classifier and scaler choices are secondary to the feature representation. Because within-session evaluation is the easiest regime, these results are a lower bound on the personalization problem: if no pipeline dominates here, none is likely to dominate under harder, participant-independent conditions. These findings suggest that a productive direction is to complement strong default pipelines with participant-aware model selection and, in time, with adaptive systems that align the feature representation to each user's neurophysiology. Confirming the pattern under explicit artefact handling and under cross-session and LOSO protocols, and extending the same statistical lens to deep and adaptive methods, are the clear next steps. To support that work, we release the full per-subject score tables and analysis code.

\section{Data and code availability}

The three datasets are publicly available through MOABB: PhysionetMI (\url{https://moabb.neurotechx.com/docs/generated/moabb.datasets.PhysionetMI.html}), Cho2017 (\ldots{}Cho2017.html), and Zhou2016 (\ldots{}Zhou2016.html). The complete per-subject score tables analysed here, and the scripts that reproduce all statistics and figures (Friedman/Nemenyi/Wilcoxon analyses, winner-per-subject and oracle computations, the dimensionality analysis, and the classifier and scaler comparisons), are provided as Supplementary Material and will be deposited in a public repository upon publication.

\section{Author contributions}

All authors contributed to the conception and interpretation of the study and to writing the manuscript. The benchmark was executed within the MOABB framework; statistical re-analysis and figures were produced from the resulting per-subject score tables.

\section{Competing interests}

O. Oullier and P. Barbaste are co-founders of Inclusive Brains, which maintains a research partnership with IBM on artificial intelligence, neurotechnologies, and quantum computing; this partnership did not influence the design, analysis, or interpretation of this study. X. Vasques is Vice-President and Chief Technology Officer of IBM Technology France, and Head of Laboratory at the Institut du Neurone. P. Barbaste is Senior Consultant at Wavestone. O. Oullier is Professor at MBZUAI, CEO of Inclusive Brains, Chairman of the Biotech Dental Group AI Institute, and a shareholder and former President of EMOTIV Inc. No other competing interests are declared.


\begin{thebibliography}{99}
\bibitem{} Khan, S. et al. Invasive brain-computer interface for communication: a scoping review. Brain Sci. 15, 336 (2025).
\bibitem{} Edelman, B. J. et al. Non-invasive brain-computer interfaces: state of the art and trends. IEEE Rev. Biomed. Eng. 18, 26--49 (2025).
\bibitem{} Saha, S. et al. Progress in brain-computer interface: challenges and opportunities. Front. Syst. Neurosci. 15, 578875 (2021).
\bibitem{} Chen, J. et al. fNIRS-EEG BCIs for motor rehabilitation: a review. Bioengineering 10, 1393 (2023).
\bibitem{} Freudenburg, Z. V. et al. Sensorimotor ECoG signal features for BCI control. Front. Neurosci. 13, 1058 (2019).
\bibitem{} Levett, J. J. et al. Invasive brain-computer interface for motor restoration in spinal cord injury: a systematic review. Neuromodulation 27, 597--603 (2024).
\bibitem{} Silva, A. B., Littlejohn, K. T., Liu, J. R., Moses, D. A. \& Chang, E. F. The speech neuroprosthesis. Nat. Rev. Neurosci. 25, 473--492 (2024).
\bibitem{} Stavisky, S. D. Restoring speech using brain-computer interfaces. Annu. Rev. Biomed. Eng. 27, 29--54 (2025).
\bibitem{} Oullier, O., Jantzen, K. J., Steinberg, F. L. \& Kelso, J. A. S. Neural substrates of real and imagined sensorimotor coordination. Cereb. Cortex 15, 975--985 (2005).
\bibitem{} Xu, L. et al. Cross-dataset variability problem in EEG decoding with deep learning. Front. Hum. Neurosci. 14, 103 (2020).
\bibitem{} Cho, H., Ahn, M., Ahn, S., Kwon, M. \& Jun, S. C. EEG datasets for motor imagery brain-computer interface. GigaScience 6, gix034 (2017).
\bibitem{} Gibson, E., Lobaugh, N. J., Joordens, S. \& McIntosh, A. R. EEG variability: task-driven or subject-driven signal of interest? NeuroImage 252, 119034 (2022).
\bibitem{} Apicella, A. et al. Toward cross-subject and cross-session generalization in EEG-based emotion recognition: systematic review, taxonomy, and methods. Neurocomputing 604, 128354 (2024).
\bibitem{} Ahn, M., Cho, H., Ahn, S. \& Jun, S. C. High theta and low alpha powers may be indicative of BCI-illiteracy in motor imagery. PLoS One 8, e80886 (2013).
\bibitem{} Becker, S., Dhindsa, K., Mousapour, L. \& Al Dabagh, Y. BCI illiteracy: it's us, not them. In 2022 10th Int. Winter Conf. on BCI 1--3 (IEEE, 2022).
\bibitem{} Kim, D.-H., Shin, D.-H. \& Kam, T.-E. Bridging the BCI illiteracy gap: a subject-to-subject semantic style transfer for EEG-based motor imagery classification. Front. Hum. Neurosci. 17, 1194751 (2023).
\bibitem{} Ramoser, H., Muller-Gerking, J. \& Pfurtscheller, G. Optimal spatial filtering of single-trial EEG during imagined hand movement. IEEE Trans. Rehabil. Eng. 8, 441--446 (2000).
\bibitem{} Barachant, A., Bonnet, S., Congedo, M. \& Jutten, C. Multiclass brain-computer interface classification by Riemannian geometry. IEEE Trans. Biomed. Eng. 59, 920--928 (2012).
\bibitem{} Barachant, A., Bonnet, S., Congedo, M. \& Jutten, C. Classification of covariance matrices using a Riemannian-based kernel for BCI applications. Neurocomputing 112, 172--178 (2013).
\bibitem{} Congedo, M., Barachant, A. \& Bhatia, R. Riemannian geometry for EEG-based brain-computer interfaces: a primer and a review. Brain-Comput. Interfaces 4, 155--174 (2017).
\bibitem{} Samek, W., Meinecke, F. C. \& Muller, K.-R. Transferring subspaces between subjects in brain-computer interfacing. IEEE Trans. Biomed. Eng. 60, 2289--2298 (2013).
\bibitem{} Singh, A. K. \& Krishnan, S. Trends in EEG signal feature extraction applications. Front. Artif. Intell. 5, 1072801 (2022).
\bibitem{} Hjorth, B. EEG analysis based on time domain properties. Electroencephalogr. Clin. Neurophysiol. 29, 306--310 (1970).
\bibitem{} Higuchi, T. Approach to an irregular time series on the basis of the fractal theory. Phys. D 31, 277--283 (1988).
\bibitem{} Astolfi, L. et al. Comparison of different cortical connectivity estimators for high-resolution EEG recordings. Hum. Brain Mapp. 28, 143--157 (2007).
\bibitem{} Islam, M. \& Lee, T. Functional connectivity analysis in multi-channel EEG for emotion detection. In Annu. Int. Conf. IEEE EMBS 1--4 (2023).
\bibitem{} Park, H. \& Jun, S. C. Connectivity study on resting-state EEG between motor-imagery BCI-literate and BCI-illiterate groups. J. Neural Eng. 21, 016015 (2024).
\bibitem{} Yger, F., Berar, M. \& Lotte, F. Riemannian approaches in brain-computer interfaces: a review. IEEE Trans. Neural Syst. Rehabil. Eng. 25, 1753--1762 (2017).
\bibitem{} Lotte, F. et al. A review of classification algorithms for EEG-based brain-computer interfaces: a 10-year update. J. Neural Eng. 15, 031005 (2018).
\bibitem{} Kunjan, S. et al. The necessity of leave-one-subject-out (LOSO) cross-validation for EEG disease diagnosis. In Brain Informatics 558--567 (Springer, 2021).
\bibitem{} Schalk, G., McFarland, D. J., Hinterberger, T., Birbaumer, N. \& Wolpaw, J. R. BCI2000: a general-purpose brain-computer interface system. IEEE Trans. Biomed. Eng. 51, 1034--1043 (2004).
\bibitem{} Goldberger, A. L. et al. PhysioBank, PhysioToolkit, and PhysioNet. Circulation 101, e215--e220 (2000).
\bibitem{} Zhou, B., Wu, X., Lv, Z., Zhang, L. \& Guo, X. A fully automated trial selection method for optimization of motor-imagery-based BCI. PLoS One 11, e0162657 (2016).
\bibitem{} Chevallier, S. et al. The largest EEG-based BCI reproducibility study for open science: the MOABB benchmark. Preprint at arXiv:2404.15319 (2024).
\bibitem{} Jayaram, V. \& Barachant, A. MOABB: trustworthy algorithm benchmarking for BCIs. J. Neural Eng. 15, 066011 (2018).
\bibitem{} Dem\v{s}ar, J. Statistical comparisons of classifiers over multiple data sets. J. Mach. Learn. Res. 7, 1--30 (2006).
\bibitem{} Friedman, M. The use of ranks to avoid the assumption of normality implicit in the analysis of variance. J. Am. Stat. Assoc. 32, 675--701 (1937).
\bibitem{} Schalk, G. et al. EEG Motor Movement/Imagery Dataset (PhysioNet). (2009).
\bibitem{} Koles, Z. J., Lazar, M. S. \& Zhou, S. Z. Spatial patterns underlying population differences in the background EEG. Brain Topogr. 2, 275--284 (1990).
\bibitem{} Barachant, A., Bonnet, S., Congedo, M. \& Jutten, C. Common spatial pattern revisited by Riemannian geometry. In 2010 IEEE Int. Workshop on Multimedia Signal Processing 472--476 (2010).
\bibitem{} Zhong, X.-C. et al. EEG-DG: a multi-source domain generalization framework for motor-imagery EEG classification. IEEE J. Biomed. Health Inform. 29, 2484--2495 (2025).
\bibitem{} Lawhern, V. J., Solon, A. J., Waytowich, N. R., Gordon, S. M., Hung, C. P. \& Lance, B. J. EEGNet: a compact convolutional neural network for EEG-based brain-computer interfaces. J. Neural Eng. 15, 056013 (2018).
\bibitem{} Schirrmeister, R. T. et al. Deep learning with convolutional neural networks for EEG decoding and visualization. Hum. Brain Mapp. 38, 5391--5420 (2017).
\bibitem{} Jayaram, V., Alamgir, M., Altun, Y., Sch\"olkopf, B. \& Grosse-Wentrup, M. Transfer learning in brain-computer interfaces. IEEE Comput. Intell. Mag. 11, 20--31 (2016).
\bibitem{} Saha, S. \& Baumert, M. Intra- and inter-subject variability in EEG-based sensorimotor brain-computer interface: a review. Front. Comput. Neurosci. 13, 87 (2020).
\bibitem{} He, H. \& Wu, D. Transfer learning for brain-computer interfaces: a Euclidean space data alignment approach. IEEE Trans. Biomed. Eng. 67, 399--410 (2020).
\end{thebibliography}
\end{document}